\documentclass[paper]{JHEP3} 

\JHEPspecialurl{http://jhep.sissa.it/JOURNAL/JHEP3.tar.gz}

\usepackage{epsfig,multicol}
\usepackage{cite}

\usepackage{amsmath}
\usepackage{amssymb}
\preprint{
KUNS-1915\\
KEK-TH-955\\
hep-th/0405096\\}

\title{Absence of a fuzzy S$^{4}$ phase
in the dimensionally reduced 5d Yang-Mills-Chern-Simons model 
}
\author{Takehiro Azuma${}^a$, Subrata Bal${}^b$, 
Keiichi Nagao${}^a$, Jun Nishimura${}^a$ \\
\llap{$^a$}Institute of Particle and Nuclear Studies, \\ 
High Energy Accelerator Research Organization (KEK),\\
1-1 Oho, Tsukuba 305-0801, Japan  \\
\llap{$^b$}Department of Physics, Kyoto University, 
Kitashirakawa,\\
Kyoto 606-8502, Japan\\
\email{azumat@post.kek.jp, 
subrata@gauge.scphys.kyoto-u.ac.jp, 
nagao@post.kek.jp, 
jnishi@post.kek.jp}} 

\abstract{
We perform nonperturbative studies of 
the dimensionally reduced 5d Yang-Mills-Chern-Simons model, 
in which a four-dimensional fuzzy manifold, ``fuzzy S$^{4}$'',
is known to exist as a classical solution.
Although the action is unbounded from
below, 
a well-defined vacuum, which stabilizes at large $N$, exists
when the coefficient of the Chern-Simons term is sufficiently small.
However, this vacuum corresponds to the ``Yang-Mills phase'',
in which the system behaves similarly to the pure Yang-Mills model.
In Monte Carlo simulations we find that
the fuzzy S$^{4}$ prepared as an initial configuration
either diverges or falls into the Yang-Mills phase.
Thus the model does not have a ``fuzzy S$^{4}$ phase''
in contrast to our previous results on the 
dimensionally reduced 3d Yang-Mills-Chern-Simons model, 
in which the ``fuzzy S$^{2}$'' can be realized dynamically.
}


\newcommand{\bel}{\begin{equation}\label}

\newcommand {\beq}{\begin{equation}}
\newcommand {\eeq}{\end{equation}}
\newcommand {\beqa}{\begin{eqnarray}}
\newcommand {\eeqa}{\end{eqnarray}}
\newcommand {\bc}{\begin{center}}
\newcommand {\ec}{\end{center}}
\newcommand {\tr}{{\rm tr\,}}

\def\vs5{\vspace*{5mm}}
\def\vs1{\vspace*{1cm}}
\def\vs2{\vspace*{2cm}}
\def\hs5{\vspace*{5mm}}
\def\hs1{\hspace*{1cm}}
\def\hs2{\hspace*{2cm}}
\def\vs50{\vspace*{50mm}}
\def\vs20{\vspace*{20mm}}

\def\tr{\hbox{tr}}

\begin{document}

\section{Introduction}
Fuzzy spheres \cite{Madore}, 
which are simple compact noncommutative manifolds,
have been discussed extensively in the literature.
One of the motivations comes from the general expectation 
that noncommutative geometry provides
a crucial link to string theory and quantum gravity.
Indeed Yang-Mills theories on noncommutative geometry
appear in a certain low-energy limit of string theory \cite{Seiberg:1999vs}.
There is also an independent observation that the space-time 
uncertainty relation, which is naturally realized by noncommutative
geometry, can be derived from some general assumptions
on the underlying theory of quantum gravity \cite{gravity}.
Another motivation is to use fuzzy spheres
as a regularization scheme alternative to 
the lattice regularization \cite{Grosse:1995ar}.
Unlike the lattice, fuzzy spheres preserve the continuous symmetries 
of the space-time considered, and hence it is expected that 
the situation concerning chiral symmetry \cite{Grosse:1994ed,
Carow-Watamura:1996wg, chiral_anomaly, non_chi,balagovi, 
chiral_anomaly2,balaGW,Nishimura:2001dq,AIN,AIN2,Ydri:2002nt, 
Iso:2002jc,Balachandran:2003ay,nagaolat03, AIN3} 
and supersymmetry might be ameliorated.

Fuzzy spheres appear as classical solutions
in matrix models with a Chern-Simons-like term
\cite{0101102,0204256,0207115,0301055}
and their dynamical properties have been studied
in refs.\ \cite{0108002,0303120,0309082,%
0312241,0401038,0402044,0403242}. 
These models belong to the class of the so-called dimensionally reduced models
(or the large-$N$ reduced models),
which is widely believed to provide a constructive definition of 
superstring and M theories \cite{9610043,9612115,9703030}.
The space-time is represented by the eigenvalues of the bosonic matrices,
and in the IIB matrix model \cite{9612115}, in particular,
the dynamical generation of {\em four}-dimensional space-time 
(in {\em ten}-dimensional type IIB superstring theory) has been discussed
by many authors \cite{Aoki:1998vn,Ambjorn:2000dx,NV,Burda:2000mn,%
Ambjorn:2001xs,exact,sign,Nishimura:2001sx,%
KKKMS,Kawai:2002ub,Vernizzi:2002mu,0307007,Nishimura:2003rj}.

In ref.\ \cite{0401038} we have studied
the dimensionally reduced 3d Yang-Mills models with the cubic 
Chern-Simons term, which has the fuzzy 2-sphere (S$^{2}$)
as a classical solution.
Unlike previous works we have performed nonperturbative first-principle
studies by Monte Carlo simulations.
We observed a first order phase transition as we vary 
the coefficient of the Chern-Simons term.
For a small coefficient
the large $N$ behavior of the model is the same as in the 
pure Yang-Mills model, whereas for a large coefficient
a single fuzzy S$^2$ appears dynamically.

For obvious reasons it is interesting to extend this work to
a matrix model which accommodates a {\em four}-dimensional fuzzy manifold.
Among various possibilities, 
here we study the dimensionally reduced 5d Yang-Mills model with the quintic
Chern-Simons term \cite{0204256,0301055}, 
which is known to have the fuzzy 4-sphere (S$^4$)
as a classical solution \cite{9712105}.
Various aspects of the fuzzy S$^4$ are discussed in refs.\
\cite{0105006,0111278,0207111,0209057,0212170,0402010,0404086}.

The rest of this paper is organized as follows. 
In section 2 we define the model
and discuss some properties of the fuzzy S$^{4}$ solution.
In section 3 we present our results of the numerical simulation.
Section 4 is devoted to a summary and discussions.
  
\section{The model and the fuzzy S$^{4}$}

The model we study is defined by the action
  \begin{eqnarray}
      S = N \left( - \frac{1}{4} \, \tr [A_{\mu}, A_{\nu}]^{2}
  - \frac{1}{5} \, \lambda \, 
         \epsilon_{\mu \nu \rho \sigma \tau } \, \tr (A_{\mu}
         A_{\nu}  A_{\rho} A_{\sigma} A_{\tau})
  \right) \ , 
\label{s4action} 
  \end{eqnarray}
where $A_{\mu}$ ($\mu = 1 , \cdots , 5$) 
are $N \times N$ traceless hermitian matrices, and 
$\epsilon_{\mu \nu  \rho \sigma \tau}$ 
is a rank-5 totally-antisymmetric
tensor with $\epsilon_{12345}=1$.

The classical equation of motion is given by
  \begin{eqnarray}
   [A_{\nu}, [A_{\mu}, A_{\nu}]] + \lambda \, 
\epsilon_{\mu \nu  \rho \sigma \tau}  \, 
A_{\nu}  A_{\rho} A_{\sigma} A_{\tau}
  = 0 \ ,
\label{s4eom}
  \end{eqnarray}
which is known to have a solution of the form
  \begin{eqnarray}
   A_{\mu} = \frac{1}{2} \, \alpha \, G_{\mu} \ , 
\label{s4sol}
  \end{eqnarray}
where $\alpha$ is given by
\begin{eqnarray}
    \alpha = \frac{1}{(n+2) \, \lambda } 
\label{s4lambda}
\end{eqnarray}
with some positive integer $n$, and $G_{\mu}$ is defined as
\beq
   G_{\mu} = \sum_{j=1}^{n}
\Bigl( 
\underbrace{{\bf 1}_{4 } \otimes \cdots \otimes {\bf 1}_{4 }}_{j-1} \otimes
\Gamma_{\mu}  \otimes \underbrace{ {\bf 1}_{4 } \otimes \cdots
       \otimes {\bf 1}_{4 }}_{n-j} 
\Bigr)_{\textrm{sym}} \ . 
\label{sympro}
\eeq
Here the tensor product $\otimes$ with the suffix ``sym'' 
represents the so-called symmetric tensor product 
(See e.g.\ ref.\ \cite{0401120} for details),
and $\Gamma_{\mu}$ are $4 \times 4$ hermitian matrices
which obey the 5-dimensional (Euclidean) Clifford algebra
  \begin{eqnarray}
   \{ \Gamma_{\mu}, \Gamma_{\nu} \} = 2 \, \delta_{\mu \nu} \ .
 \label{clifford}
  \end{eqnarray}
An explicit representation 
of the 5d gamma matrices $\Gamma_{\mu}$ is
\beqa
   \Gamma_{a} &=&  \sigma_{a} \otimes \sigma_{1} \quad \quad (a=1,2,3) \ , \\
   \Gamma_{4} &=& {\bf 1}_{2 } \otimes \sigma_{2}  \ , \\
   \Gamma_{5} &=& {\bf 1}_{2 } \otimes \sigma_{3} \ ,
  \label{gammadef}
\eeqa
where $\sigma_{a}$ ($a=1,2,3$) are the Pauli matrices .
According to the definition of the symmetric tensor product,
the size $N$ of the matrix $G_{\mu}$ in (\ref{sympro}) is given by
  \begin{eqnarray}
   N = \frac{1}{6} \, (n+1)(n+2)(n+3)   \ . 
\label{s4matrixsize}
  \end{eqnarray}
The solution (\ref{s4sol}) can be regarded as the fuzzy S$^{4}$
\cite{0204256,0301055}.
\footnote{More precisely,
the solution (\ref{s4sol}) corresponds to the   
homogeneous space SO$(5)$/U$(2)$, which is locally S$^4\times$S$^2$.
Here the fuzzy S$^{2}$ is naturally considered as an 
internal space attached to each point on the fuzzy S$^{4}$
\cite{0204256,0301055}.
}
Unlike the fuzzy S$^{2}$ case \cite{0401038}, 
the fuzzy S$^{4}$ solution exists
only for particular values of $N$ given by (\ref{s4matrixsize}).

{}From the algebraic point of view, 
the fuzzy S$^{4}$ solution (\ref{s4sol}) differs from
the fuzzy S$^{2}$ in that
$G_{\mu}$ is not closed with respect to the commutator.
Let us therefore define the commutator as
$G_{\mu \nu} = [G_{\mu}, G_{\nu}]$. 
Then $G_{\mu}$ and $G_{\mu\nu}$ satisfy the following relations 
\cite{9712105,0105006,0401120}
\beqa
G_{\mu} G_{\mu} &=& n \, (n+4) \, {\bf 1}_{N } \ , \label{s4prop1}  \\
\label{s4prop2} 
G_{\mu \nu} G_{\mu \nu} &=& - 16 \, n \, (n+4) \, {\bf 1}_{N } \ , \\
\label{s4prop3}
[ G_{\mu\nu}, G_{\rho} ] &=& (- \delta_{\mu \rho} G_{\nu} +
  \delta_{\nu \rho} G_{\mu}) \ ,  \\
\label{s4prop4}
[ G_{\mu \nu}, G_{\rho \chi} ] &=& 
(\delta_{\nu \rho} G_{\mu \chi} + \delta_{\mu \chi} 
G_{\nu \rho} - \delta_{\mu \rho} G_{\nu \chi} 
- \delta_{\nu \chi} G_{\mu\rho}) \ . 
\eeqa
{}From eq.\ (\ref{s4prop1}) we obtain
\beqa
\label{s4radius}
   \sum_{\mu}(A_{\mu})^{2} 
&=& R^2 \,  {\bf 1}_{N } \ , \\
R^2 &=& \frac{1}{4} \, \alpha^{2} \, n \,  (n+4)  \ , 
\label{defR}
  \end{eqnarray}
which means that 
$R$ represents
the ``radius'' of the fuzzy S$^4$.
Note that the radius $R$ is {\em inversely} proportional to $\lambda$,
the coefficient of the Chern-Simons term, unlike the fuzzy S$^2$ case
\cite{0401038}.
The matrix $G_{\mu}$ also satisfies the self-duality condition
  \begin{eqnarray}
   \epsilon_{\mu \nu \rho \sigma \tau } G_{\nu}
   G_{\rho} G_{\sigma} G_{\tau}
  = 8 \, (n+2) \, G_{\mu} \ . \label{s4selfduality}
  \end{eqnarray}
Using these formulae, it is easy to check that (\ref{s4sol})
is a solution to the classical equation of motion (\ref{s4eom}).
The value of the action for the classical solution (\ref{s4sol}) is
given as
   \begin{eqnarray}
    S_{\rm cl} = \frac{1}{20} \,  n \, (n+4) \, \alpha^{4} \, N^{2} \ ,  
\label{s4classicalenergy}
   \end{eqnarray}
which {\em increases} with $\alpha$ and
hence with the radius $R$ of the fuzzy S$^4$.
This is again in contrast to the fuzzy S$^{2}$ case \cite{0401038}, 
where the corresponding action has an overall minus sign,
and therefore it {\em decreases} with the radius of the fuzzy S$^2$.

In order to reveal some crucial properties of the present model,
let us consider a one-parameter family of configurations
\beq
A_\mu = \frac{1}{2} \, \xi \, G_\mu \ ,
\label{rescaled}
\eeq
where $\xi = \alpha$ corresponds to the classical solution (\ref{s4sol}).
The corresponding value of the action (\ref{s4action})
for each $\xi$ is given by
\beq
S(\xi) = N^2 \, n \, (n+4) \left( 
\frac{1}{4} \xi^{4} \, - \frac{1}{5 \, \alpha} \xi^{5} \right)  \ .
\label{Sbeta}
\eeq
We find that the action can take arbitrarily negative value
by increasing $\xi$.
Thus the classical action is not bounded from below
and the path integral over $A_\mu$ obviously diverges. 
Moreover, the point $\xi = \alpha$ actually gives 
a local maximum of $S(\xi)$.
Thus the solution (\ref{s4sol}) is merely a saddle point
of the classical action.
These properties are in striking contrast to the situation
in the fuzzy S$^2$ case \cite{0401038}, where the single fuzzy S$^2$
is a minimum of the action (at least locally), 
and the path integral converges for $N \ge 4$ \cite{0310170}.
These differences come from the fact that the Chern-Simons term
in the present model is {\em quintic}
and it has a higher power in $A_\mu$ than the Yang-Mills term,
whereas in the previous model the Chern-Simons term is {\em cubic}
and it has a lower power in $A_\mu$ than the Yang-Mills term.

Even if the path integral diverges for $\lambda \neq 0$ at finite $N$, 
there is a possibility that
the model possesses a well-defined vacuum, which stabilizes at
large $N$. A well-known example is the one-matrix model
\begin{eqnarray}
   S_{\phi} 
= N \left( \frac{1}{2} \, \tr \, \phi^{2} 
- \frac{1}{3} \, g \,  \tr \, \phi^{3}
\right), \label{onematrix}
\end{eqnarray}
where $\phi$ is an $N \times N$ hermitian matrix. 
Although the action is unbounded from below, 
this model has a sensible large-$N$ solution
\cite{Brezin:1977sv} for $g^2 \leq g_c^2 \equiv \frac{1}{12\sqrt{3}}$.
%
Similarly, our model (\ref{s4action}) does possess a well-defined vacuum
for sufficiently small $\lambda$, as we will see in the next section.
We will also examine the possibility that
the fuzzy S$^4$, which is unstable classically,
may get stabilized due to some quantum effects.

  \section{Monte Carlo simulations}
We address the issues raised in the previous section
by Monte Carlo simulation.
For an updating procedure, we apply the heat bath algorithm 
developed in ref.\ \cite{9811220}
(See the appendix A of ref.\ \cite{0401038} for more details).
  We take the fuzzy S$^4$ solution (\ref{s4sol}) as the initial 
  configuration, and carry out simulations for various $\lambda$
  at $N=4,10,20,35,56$, which corresponds to $n=1,2,3,4,5$, respectively, 
  due to (\ref{s4matrixsize}).
Since the qualitative behavior turns out to be the same for any $N$,
we will show explicit results only for $N=56$.
  In order to probe the geometrical structure of the configurations,
  we consider the eigenvalues $\lambda_{j}$ ($j=1, \cdots , N$) of 
  the quadratic Casimir operator \cite{0401038}
   \begin{eqnarray}
    Q = \sum_{\mu=1}^{5} (A_{\mu})^{2} \ .
    \label{s2casimir}
   \end{eqnarray}

  In figure \ref{a04}
  we plot the eigenvalues $\{ \lambda_{j} \}$
  and the normalized action $\frac{S}{N^{2}}$ 
  for $\alpha = 0.4$ 
  against the Monte Carlo time $\tau$ 
  (or more precisely, the number of ``sweeps'' in 
  the heat bath algorithm \cite{0401038}).
  For the initial fuzzy S$^4$ configuration ($\tau=0$), 
  all the eigenvalues coincide with (\ref{defR})
  and the action is given by (\ref{s4classicalenergy}).
  After the first sweep, the eigenvalues spread out, 
  obviously for entropical reasons, and correspondingly the action increases.
  The eigenvalues and the action are stable for several sweeps, 
  but change drastically at $\tau = 6 $.
  At $\tau = 7$, which is not included in the figure,
  the maximum eigenvalue reaches $2.0\times 10^{4}$
  and the action becomes $\frac{S}{N^2}=-1.7\times 10^{6}$.
  Subsequently all the eigenvalues blow up and
  the action takes further negative values until the run terminates due to
  overflow. 
  We observed similar behaviors for $\alpha \lesssim 0.5$ 
  (i.e., large $\lambda$).

     \FIGURE{
    \epsfig{file=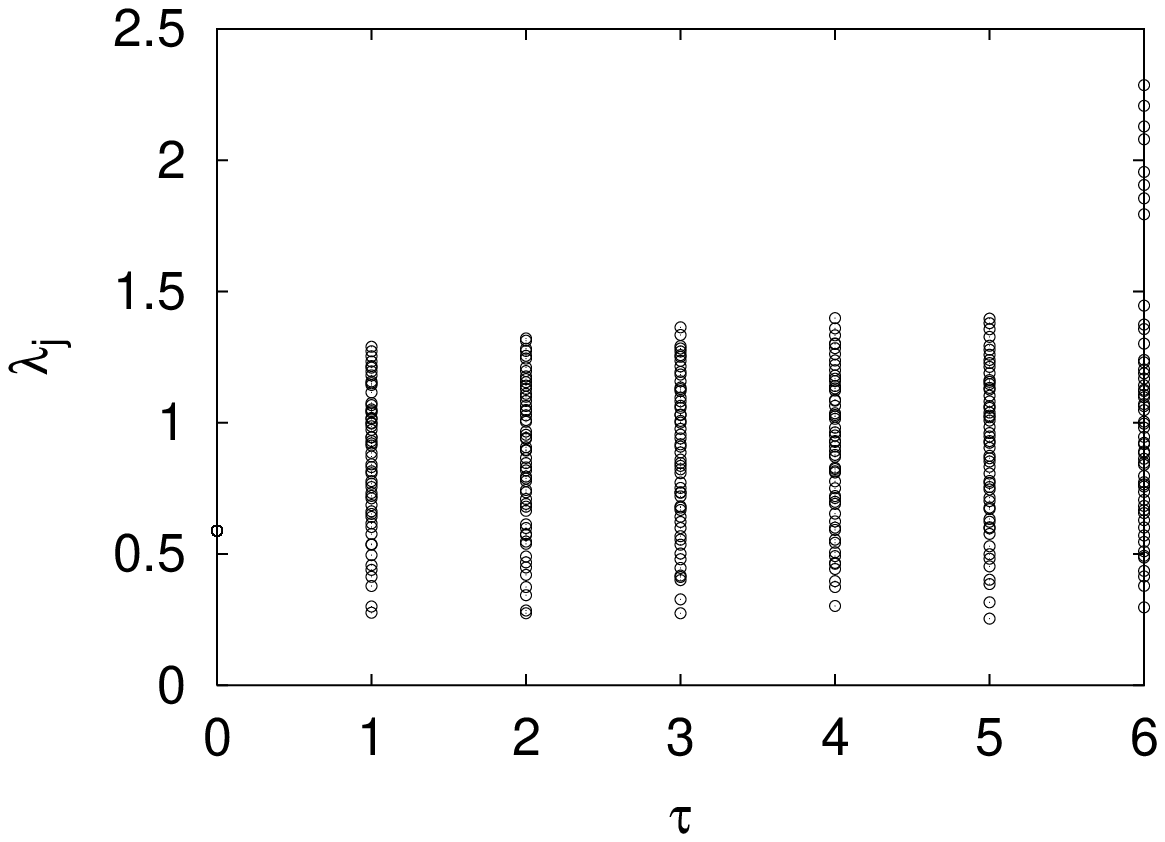,width=7.4cm}
    \epsfig{file=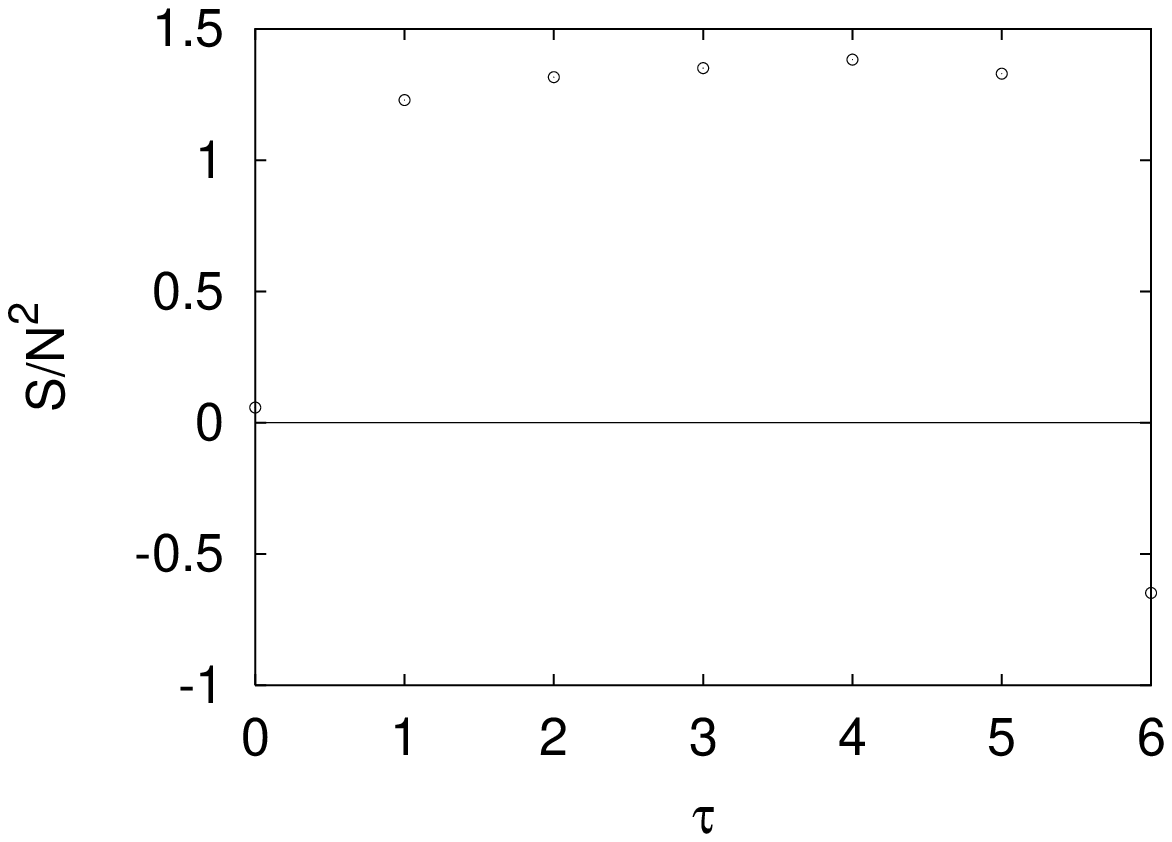,width=7.4cm}
   \caption{The eigenvalues $\{ \lambda_{j} \}$ of the quadratic Casimir 
   operator $Q$ (left) 
   and the normalized action $\frac{S}{N^{2}}$ (right)
   are plotted against the Monte Carlo time $\tau$ 
   for $\alpha = 0.4$ at $N=56$.}
   \label{a04}}

  For $\alpha \gtrsim 0.5$ (i.e., small $\lambda$), on the other hand, 
  the system turns out to be well-behaved.
  In figure \ref{s4a06} we plot the eigenvalues $\{ \lambda_{j} \}$
  and the normalized action $\frac{S}{N^{2}}$
  against the Monte Carlo time $\tau$.
  The eigenvalues spread out after the first sweep, and then they all 
  decrease keeping their extent almost constant. The action goes up
  after the first sweep, 
  and then goes down as all the eigenvalues decrease.
  This behavior can be understood qualitatively 
  from the action (\ref{Sbeta}) for the
  rescaled configuration (\ref{rescaled}).
  The action after thermalization turns out to be 
  almost independent of $\alpha$.
  This is a property of the ``Yang-Mills phase''\cite{0401038}, in which
  the effect of the Chern-Simons term is negligible and the model
  behaves similarly to the pure Yang-Mills model ($\lambda = 0$, i.e.,
  $\alpha = \infty$ )
  \cite{9811220}.

    \FIGURE{
    \epsfig{file=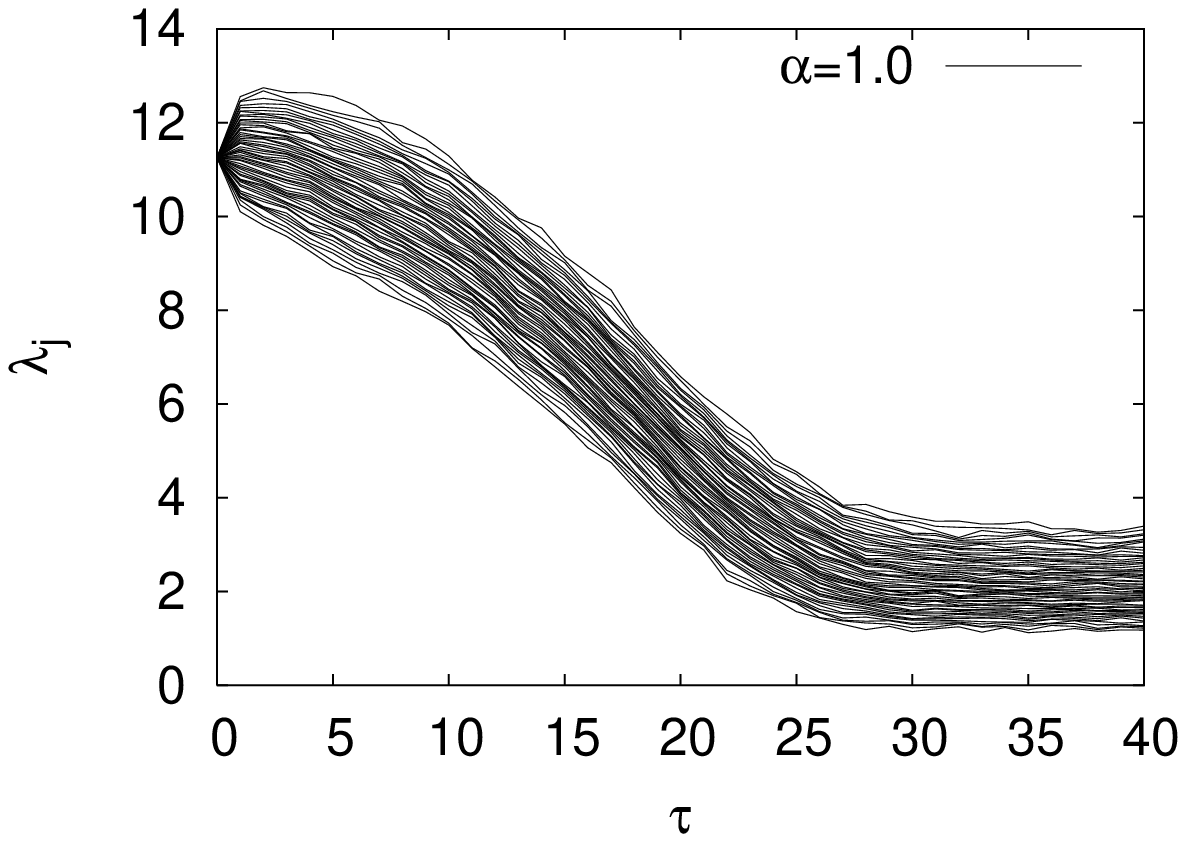,width=7.4cm}
    \epsfig{file=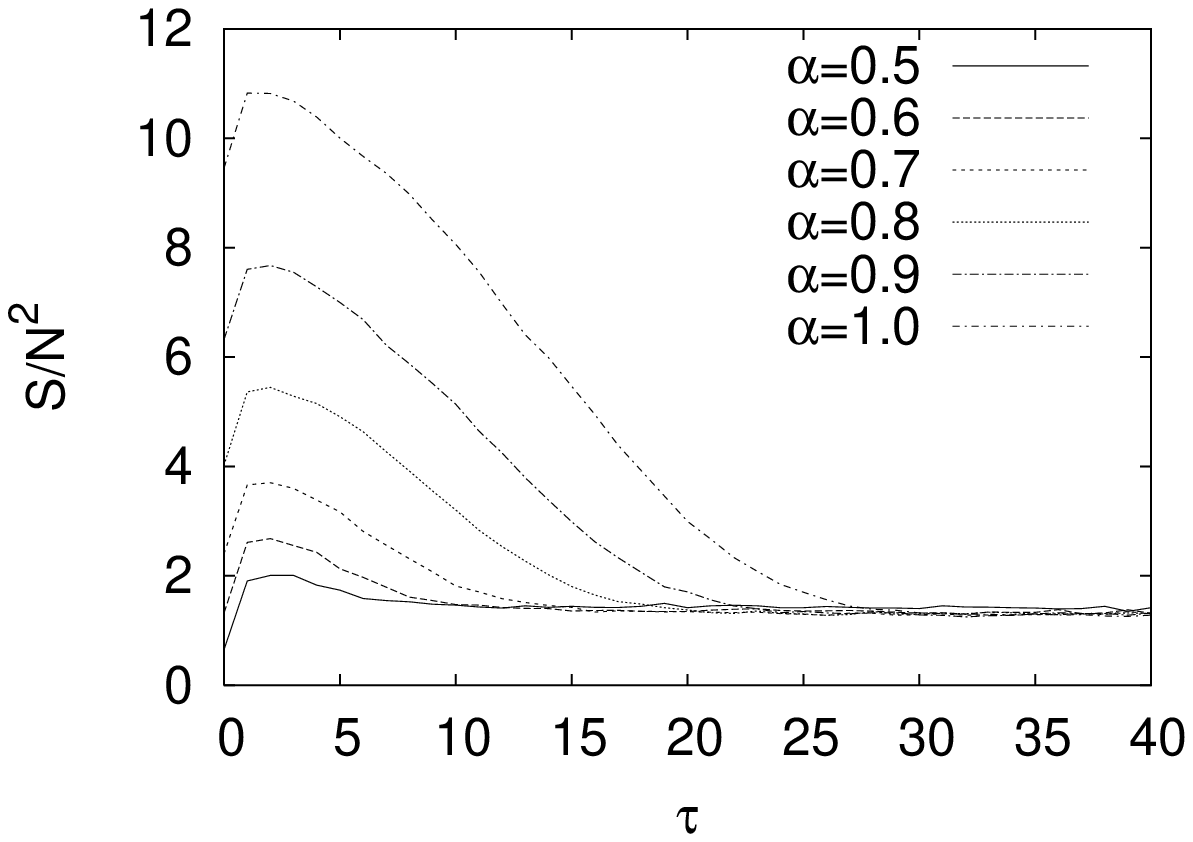,width=7.4cm}
   \caption{The 56 eigenvalues of the quadratic Casimir operator 
   $Q$ for $\alpha =1.0$ (left) and
   the normalized action $\frac{S}{N^{2}}$
   for $\alpha = 0.5, \cdots,1.0$ (right)
   are plotted against the Monte Carlo time $\tau$ at $N=56$.}
   \label{s4a06}}

    \FIGURE{
    \epsfig{file=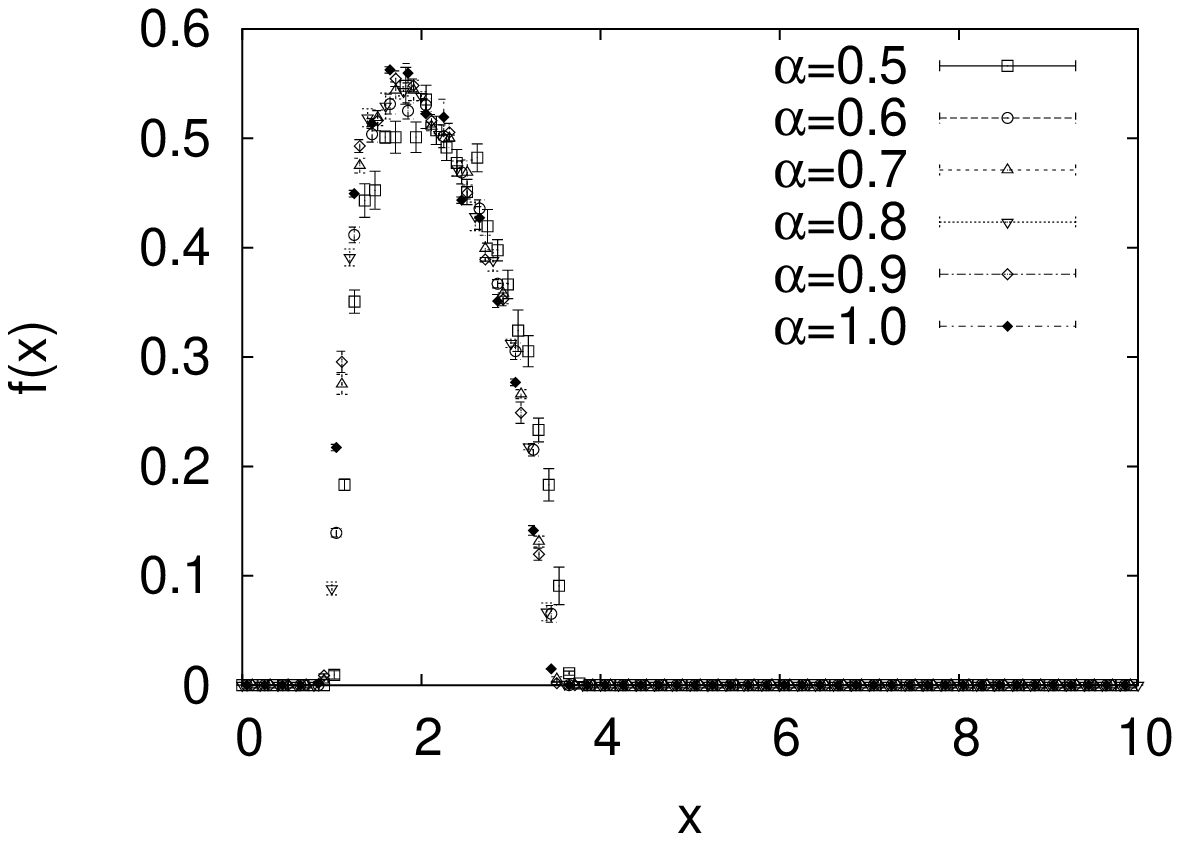,width=7.4cm}
   \caption{The eigenvalue distribution $f(x)$ is plotted 
    for $\alpha = 0.5, \cdots, 1.0$ at $N=56$.}
   \label{histn}
    }

  We also calculate the eigenvalue distribution defined by
   \begin{eqnarray}
     f(x) = \frac{1}{N} \sum_{j=1}^{N} \langle \delta(x-\lambda_{j}) 
     \rangle \ .
   \end{eqnarray}
   For the fuzzy S$^4$ solution, it is given within the classical 
  approximation as
\beq
f(x) = \delta (x - R^2 ) \ ,
\eeq
where $R^2$ defined by (\ref{defR}) is proportional to $\alpha^2$.
Figure \ref{histn} shows the results for $\alpha = 0.5, \cdots, 1.0$.
The results are almost independent of $\alpha$, which continues
to be the case for even larger $\alpha$.
This is again a property of the ``Yang-Mills phase''.
%
Since the empty region 
at $x \lesssim 1$
can be understood from the uncertainty principle,
the geometry should rather be considered as 
that of a solid ball \cite{0401038}.

We have also simulated the model using $A_\mu = 0$ 
as the initial configuration.
In this case we are not restricted to the values of 
$N$ given by (\ref{s4matrixsize}), so
we tried $N=12,14,16$ as well.
For any value of $N$, we find that 
the system either diverges or falls into the Yang-Mills phase.
   
\section{Summary and discussions}
  In this paper we reported on a first attempt to 
  extend our previous work \cite{0401038}
  on the fuzzy S$^2$ to a four-dimensional fuzzy manifold.
  We considered the fuzzy S$^4$ realized as a classical solution
  in the dimensionally reduced 5d Yang-Mills-Chern-Simons model.
  We find various differences between the present model
  and the previous one in spite of their similarity.
  In particular we have observed that the fuzzy S$^{4}$ prepared 
  as the initial configuration either diverges
  or falls into the Yang-Mills phase.
  Thus we conclude that the present model does not have a phase 
  in which the fuzzy S$^{4}$ becomes stable.


  This negative result
  is essentially due to the fact that 
  the quintic Chern-Simons
  term has higher powers in $A_\mu$ than the Yang-Mills term.
\footnote{Fuzzy spheres appear as classical solutions also in dimensionally
reduced Yang-Mills models with a tachyonic mass term \cite{0103192}.
We obtained clear indications from Monte Carlo simulations 
that such systems are ill-defined.} 
  In this regard we note that there are other models 
  which accommodate four-dimensional fuzzy manifolds 
  such as CP$^2$ and ${\rm S}^2\times
  {\rm S}^2$ with some kind of a
  {\em cubic} Chern-Simons-like term. 
  We would like to report on these cases in
  the forthcoming publications.

\acknowledgments
We would like to thank Hikaru Kawai and Yusuke Kimura for helpful discussions. 
The work of T.A., S.B.\ and J.N.\ 
is supported in part by Grant-in-Aid for 
Scientific Research (Nos.\ 03740, P02040 and 14740163, respectively)
from the Ministry of Education, Culture, Sports, Science and Technology.

\end{document}